\documentclass[conference]{IEEEtran}
\IEEEoverridecommandlockouts

\usepackage{cite}
\usepackage{amsmath,amssymb,amsfonts}
\usepackage{algorithmicx}
\usepackage{graphicx}
\usepackage{textcomp}

\usepackage{algorithm}
\usepackage{algpseudocode}
\usepackage{booktabs}
\usepackage{pgfplots}
\usepackage{multirow}
\usepackage{url}
\usepackage{tikz}
\usetikzlibrary{positioning,arrows.meta,shapes.geometric,calc}
\usepackage{hyperref}
\usepackage{xspace}
\usepackage{bm}
\usepackage[table]{xcolor}
\usepackage{threeparttable}

\newcommand{\ghost}{\textsc{Ghost}\xspace}

\def\BibTeX{{\rm B\kern-.05em{\sc i\kern-.025em b}\kern-.08em
    T\kern-.1667em\lower.7ex\hbox{E}\kern-.125emX}}
\begin{document}

\title{\ghost: Plausible Yet Unlearnable Trajectories via On-Manifold Substitution for Next-POI Privacy}

\author{\IEEEauthorblockN{
Zhenyu Yu\textsuperscript{1}, 
Mohd Yamani Idna Idris\textsuperscript{2}, 
Jihong Guan\textsuperscript{3}, 
Shuigeng Zhou\textsuperscript{1}}
\IEEEauthorblockA{
\textit{
\textsuperscript{1}Fudan University, 
\textsuperscript{2}University of Malaya, 
\textsuperscript{3}Tongji University}\\
\texttt{yuzhenyuyxl@foxmail.com, 
yamani@um.edu.my, 
jhguan@tongji.edu.cn, 
sgzhou@fudan.edu.cn}}
}

\maketitle

\begin{abstract}
A publisher who releases check-in trajectories inadvertently publishes a strong predictor of every user's future locations. We address this risk by generating \emph{unlearnable trajectories}, perturbed sequences that yield victim models with degraded next-Point-of-Interest (next-POI) accuracy on clean test inputs. Direct ports of image-domain unlearnable examples fail on two counts. The published data must remain geographically and semantically plausible, and the perturbation must resist purification adversaries that exploit the structure of randomized defences. We propose \ghost, a manifold-aligned framework whose perturbations look like plausible human check-in sequences yet leave no learnable signal behind. \ghost steers each substitution onto the real-trajectory manifold through a frozen trajectory language model, so a denoising-bridge adversary has nothing to invert and a context-free frequency-table adversary recovers a near-uniform distribution. Across two standard benchmarks, and four attacker postures, \ghost achieves protection-gap competitive with the strongest deterministic baseline (PGD) while attaining the lowest restored accuracy under the bigram adaptive purification adversary on both datasets, and lies within one per-cell standard deviation of PGD on the protection-versus-purification-resistance plane. Ablations confirm the manifold prior subsumes the entropy-floor knob of prior randomized defences, with the frequency-table adversary's survival gap remaining within $0.04$ even when twenty percent of the pairs are leaked.
\end{abstract}

\begin{IEEEkeywords}
Unlearnable Examples, Trajectory Privacy, Next-POI Prediction, Purification Attack, Manifold Learning.
\end{IEEEkeywords}

\begin{figure}
    \centering
    \includegraphics[width=1\linewidth]{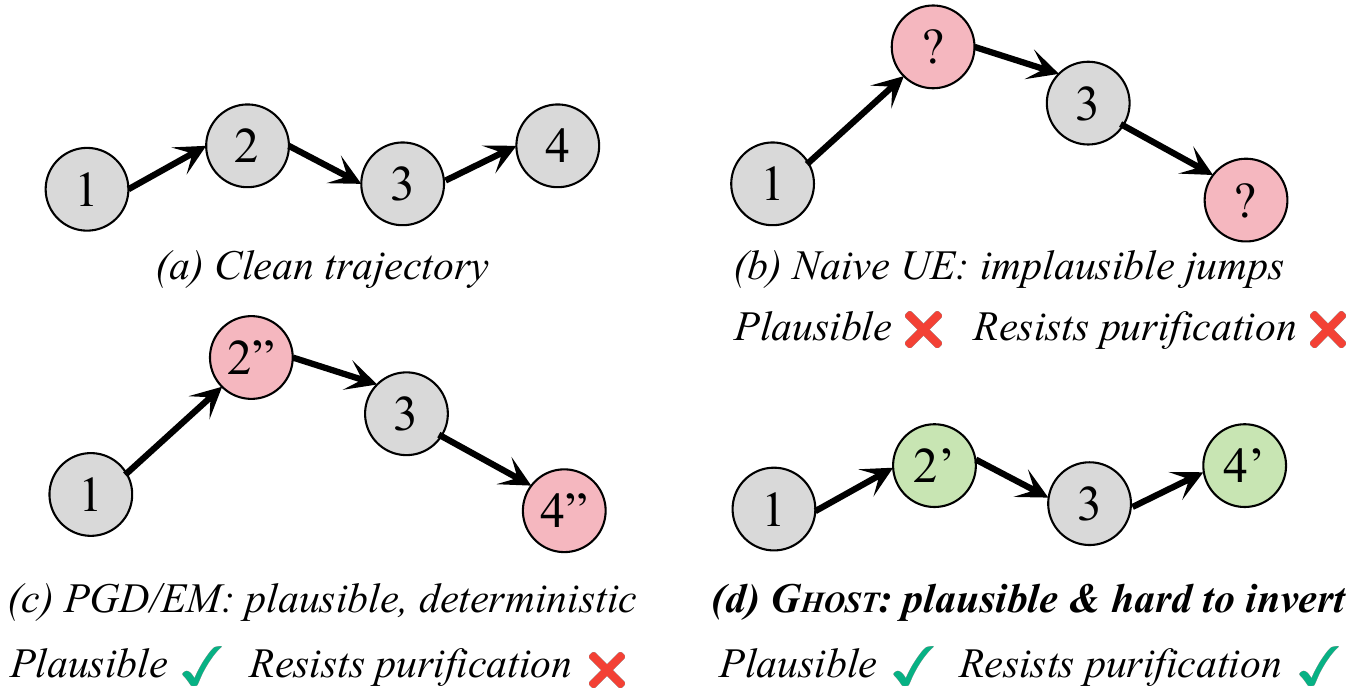}
    \caption{Conceptual comparison of trajectory perturbation strategies. (a) A user's clean POI trajectory. (b) Naive UE substitutes ignore geographic and category plausibility, producing implausible jumps. (c) Plausibility-aware deterministic baselines (PGD, EM) lie on the real-trajectory manifold but produce a fixed map that purification adversaries can learn to invert. (d) \ghost keeps each substitution on the manifold and samples stochastically from the high-density region of the prior, simultaneously preserving plausibility and resisting purification.}
    \label{fig:concept}
\end{figure}

\section{Introduction}
\label{sec:intro}

Releasing check-in trajectories is, by construction, releasing a strong predictor of every contributor's future whereabouts. Modern next-Point-of-Interest (next-POI) models~\cite{luo2021stan,yang2022getnext} trained on public datasets such as Foursquare TSMC2014~\cite{yang2015modeling} and Gowalla~\cite{cho2011friendship} recover individual-level next-location predictions with accuracy that supports commercial profiling, surveillance, and harassment far beyond the original research purpose. The temporal regularity that makes these datasets scientifically valuable also makes them privately dangerous, and the danger materializes the moment the data leave the publisher's hands. A defence is therefore needed at \emph{release} time, in a form that destroys the data's value as supervised training signal while preserving its statistical value to downstream services.

Withholding the data is not an option, since public check-in datasets are the substrate for academic next-POI research, urban-planning collaborations, and reproducibility of dozens of published models. Machine unlearning~\cite{liu2025threats} also fails to fit, since it modifies a trained model post hoc to remove specific data influence but presupposes a cooperative model owner, whereas in the public-release setting any third party can train an arbitrary model. The publisher therefore needs a release-time mechanism that irreversibly suppresses the unwanted downstream capability while preserving the wanted ones.

The unlearnable examples (UE) paradigm~\cite{huang2021unlearnable,zhu2024detection,liu2024multimodal,yu2024unlearnable}, with subsequent extensions for adversarial robustness~\cite{fu2022robust}, architecture transferability~\cite{ren2023transferable}, and sequential data~\cite{jiang2024unlearnable}, offers a natural template: the publisher releases a perturbed dataset such that any victim model trained on it fails to generalize to clean inputs. Porting UE from images to POI trajectories, however, exposes two domain-specific failures of prior work. \textbf{Plausibility.} The discrete analog of image-domain noise is POI substitution, and any unconstrained substitution can produce trajectories that are geographically impossible, semantically absurd, or speed-implausible, breaking downstream services and alerting auditors to the manipulation. \textbf{Purification adversaries.} Recent attacks invert the publisher's perturbation map. BridgePure~\cite{wang2024bridgepure} trains a Schr\"odinger-bridge denoiser from a few leaked (clean, protected) pairs and restores most of the original signal at 5\% leakage, and a simpler frequency-table adversary, in the spirit of adversarial fine-tuning~\cite{tao2021better}, builds the empirical conditional $\hat{p}(\text{clean} \mid \text{perturbed})$ and applies its argmax. Both exploit the predictable structure that randomized defences such as the entropy-floor randomization of error-minimizing UE inevitably leave behind.

We propose \textbf{\ghost}, a bilevel unlearnable-trajectories framework whose central idea is to replace entropy-floor randomization with a \emph{manifold prior}. Each released trajectory is a ghost of the user's real one: \emph{visible} as a plausible check-in sequence and \emph{intangible} as training signal. \ghost scores each candidate substitution as a linear combination of an adversarial term that is error-maximizing against a surrogate victim and a manifold term equal to the log-likelihood of the candidate under a frozen trajectory language model trained on real check-in sequences. Hard plausibility constraints on the candidate set address the plausibility failure. The manifold term addresses the purification failure structurally: every published substitution is supported by the real-trajectory distribution, so the optimal denoiser on \ghost's output approaches the identity and the optimal frequency-table inverter approaches uniform over the candidate set. The same prior that guarantees plausibility is what makes purification fail. \ghost lies on the protection-vs-purification-resistance Pareto frontier on both datasets and is the strongest defender against the bigram adaptive frequency-table attack among all high-protection methods. \ghost is contrasted with both unconstrained and plausibility-aware baselines in Figure~\ref{fig:concept}.

\textbf{Contributions.}
\begin{itemize}
\item \textbf{Threat model and plausibility.} We formalize unlearnable trajectories under hard plausibility constraints and a three-attacker purification threat model covering denoising-bridge inversion, frequency-table inversion, and bigram-adaptive purification.

\item \textbf{Manifold prior.} We propose \ghost, whose substitution score combines an adversarial term with a manifold-prior term from a frozen trajectory language model. The prior replaces entropy-floor randomization and yields a structural argument for purification resistance.

\item \textbf{Pareto-frontier protection.} \ghost\ matches or beats three baselines across four attacker postures, attaining the lowest restored accuracy under the frequency-table inverter on NYC and under the bigram-adaptive purifier on both datasets, with graceful degradation up to twenty percent adversary leakage.
\end{itemize}

\section{Related Work}
\label{sec:related}

\subsection{Unlearnable Data}
The UE paradigm was introduced by Huang et al.~\cite{huang2021unlearnable}, who showed that adding small error-minimizing noise to image training data can make the resulting classifier fail to generalize. The optimization is bilevel, with an inner surrogate fit on the perturbed data and an outer perturbation that drives the surrogate's training loss toward zero, making the data look already-learned and preventing genuine feature acquisition. Subsequent work refined this template. Robust UE~\cite{fu2022robust} targets adversarially-trained victims that ignore the original error-minimizing noise, and Transferable UE~\cite{ren2023transferable} reduces dependence on a fixed surrogate architecture. All of these methods operate on continuous pixel spaces. The closest sequential analog is a recent time-series adaptation~\cite{jiang2024unlearnable}, which we use as a baseline. No prior UE work treats the discrete, plausibility-constrained POI setting we address.

\subsection{Trajectory and POI Privacy}
Trajectory privacy has a long history grounded in $k$-anonymity and differential privacy. Never-Walk-Alone~\cite{abul2008never} anonymizes moving-object databases through uncertainty, and DPT~\cite{he2015dpt} synthesizes differentially-private trajectory samples via hierarchical reference systems. These methods provide formal privacy with respect to identifiability but do not target the unlearnable-data threat model. A downstream learner can often extract useful next-POI signal from $k$-anonymized or DP-synthesized data, especially when the synthesis preserves marginal distributions. We position \ghost as the first plausibility-and-purification-aware UE framework for next-POI trajectories.

\subsection{Purification Attacks and Defences}
Purification attacks target defences that add predictable noise to the released data. BridgePure~\cite{wang2024bridgepure} trains a Schr\"odinger-bridge model from a few leaked (clean, protected) pairs and uses it as an inference-time sampler that inverts the protection, recovering most of the clean signal at only $5\%$ leakage on image-domain UE. A simpler complementary adversary, which we instantiate, treats the perturbation map as a labelled lookup table by building $\hat{p}(\text{clean} \mid \text{perturbed})$ from leaked pairs and applying its argmax. Adversarial fine-tuning~\cite{tao2021better} is a complementary line in which the attacker, rather than restoring the data, trains a victim with augmentation that exploits the assumed noise structure. Each of these attacks shares a common weakness against a defence whose perturbation map is supported on a high-density region of the data manifold. The denoiser finds nothing unexpected to remove, the frequency table degenerates to a near-uniform distribution because every plausible candidate appears with comparable frequency, and the fine-tuning augmentation cannot model noise that is not there. \ghost is designed around this observation.

\subsection{Relation to Machine Unlearning}
\ghost shares a high-level goal with the \emph{machine unlearning} literature~\cite{liu2025threats}, namely removing certain training-data influence from downstream models, but operates at the opposite end of the pipeline. Machine unlearning modifies a trained model $\theta$ post hoc so as to approximate a model that never saw a specific subset of samples, and is invoked reactively when regulatory or user requests demand removal of already-absorbed information. Recent results show that unlearning algorithms struggle to suppress poisoning influence once it has been absorbed~\cite{pawelczyk2025machine}, and similar fragility appears in federated unlearning~\cite{wang2025poisoning}. \ghost takes the complementary, proactive route. By perturbing the released data so that no useful signal is absorbed in the first place, \ghost prevents the very condition that machine unlearning is designed to remediate. The two paradigms are therefore complementary rather than competing, and a defence-in-depth deployment could combine on-release \ghost protection with on-demand unlearning for residual cases.

\section{Threat Model and Preliminaries}
\label{sec:threat}

\subsection{Notation}
Let $\mathcal{P}$ be a finite POI vocabulary with $|\mathcal{P}| = P$. A user check-in session is an ordered sequence $s = (p_1, t_1, p_2, t_2, \dots, p_L, t_L)$ with $p_i \in \mathcal{P}$ the visited POI and $t_i \in \mathbb{Z}_{\ge 0}$ a unix timestamp. We write $s_{\le i}$ for the prefix $(p_1, t_1, \dots, p_i, t_i)$ and $s_{<i}$ for the strict prefix. A dataset is a set of sessions $\mathcal{D} = \{s_u\}_{u \in \mathcal{U}}$. We reserve $p_{\text{pad}} \in \mathcal{P}$ (index $0$) as a padding token, with $p_{\text{pad}} \notin C(p)$ for all $p \in \mathcal{P}$.

\subsection{Threat Model}
\label{sec:roles}
We model three roles. The \emph{publisher} holds the clean dataset $\mathcal{D}_{\text{clean}}$ and releases a perturbed version $\mathcal{D}_{\text{prot}} = \{M(s_u)\}$, where the per-session map $M$ substitutes some positions with alternative POIs from a plausibility-filtered candidate set $C(p)$. The publisher maximizes the gap between victim accuracy on $\mathcal{D}_{\text{clean}}$ and on $\mathcal{D}_{\text{prot}}$, subject to $M$ respecting the plausibility constraints in Section~\ref{sec:plaus}. The \emph{victim} trains a next-POI prediction model on whatever data is available, instantiated as a two-layer causal Transformer following STAN~\cite{luo2021stan}. The \emph{adversary} sits between the publisher's release and the victim, and tries to recover usable training signal from $\mathcal{D}_{\text{prot}}$. We evaluate one no-adversary baseline (A1) and three complementary purification adversaries (A2, A3, and A4) that differ in expressive power and structural assumptions.

\textbf{(A1) No adversary.}\quad The victim trains directly on $\mathcal{D}_{\text{prot}}$. This is the publisher's best case and the standard UE evaluation setting.

\textbf{(A2) Denoising-bridge purifier.}\quad The adversary obtains a fraction $r$ (default $r = 0.05$) of (clean, perturbed) pairs from the release pipeline through caching, archived data, or insider access, matching the threat model of BridgePure~\cite{wang2024bridgepure}. It trains a sequence denoiser $\hat{M}^{-1}_\theta$ on the leaked pairs and applies it iteratively to the non-leaked perturbed sessions. The victim then trains on the restored dataset.

\textbf{(A3) Frequency-table inverter.}\quad The adversary has the same leaked pairs as A2 but uses them differently. It builds the empirical context-free conditional distribution
\begin{equation}
\hat{p}(p^{\text{clean}} \mid p^{\text{prot}}) = \frac{
  \#\{(p^{\text{clean}}, p^{\text{prot}}) \text{ in leaked pairs}\}
}{\#\{p^{\text{prot}} \text{ in leaked pairs}\}}
\end{equation}
and replaces every perturbed position by its argmax under $\hat{p}$, falling back to the original perturbed POI for unseen keys. A3 is weaker than A2 in expressive power but provides a sanity check. A defence broken by A3 is decisively broken, while a defence broken by A2 but not A3 indicates the perturbation map is non-bijective, which is desirable.

\textbf{(A4) Bigram-adaptive purifier.}\quad The adversary has the same leaked pairs as A2 and A3 but builds a bigram conditional table $\hat p(p^{\text{clean}}_i \mid p^{\text{prot}}_i, p^{\text{prot}}_{i-1})$ that exploits one step of perturbed context, falling back to the unigram conditional of A3 when the bigram key is unseen. A4 is an adaptive strengthening of A3 and tests whether sequence-level structure leaks under the perturbation map.

\subsection{Plausibility Constraints}
\label{sec:plaus}
For the released data to be deployable in downstream services, every substitution $p \mapsto p'$ used by any method we evaluate must satisfy three plausibility constraints. \textbf{Geographic plausibility} requires $d_{\text{hav}}(p, p') \le R$ with $R = 1.0\,\text{km}$ (haversine distance), a typical urban-mobility radius that accommodates walking-distance alternatives while excluding cross-district teleportation. \textbf{Semantic plausibility} requires $\mathrm{category}(p') = \mathrm{category}(p)$, a hard constraint that preserves the visited POI's functional type. \textbf{Speed plausibility} bounds the implied speed between consecutive perturbed POIs by 60\,km/h, a conservative upper bound that accommodates all common urban modalities (walking, transit, driving on city streets) while excluding cross-city teleportation. These constraints define the candidate set $C(p) \subseteq \mathcal{P}$ that every protection method draws from in our experiments. The candidate set is therefore not a contribution specific to \ghost but a shared evaluation harness ensuring that algorithmic differences are not confounded by differences in plausibility filtering.

\begin{figure*}[t]
    \centering
    \includegraphics[width=0.99\linewidth]{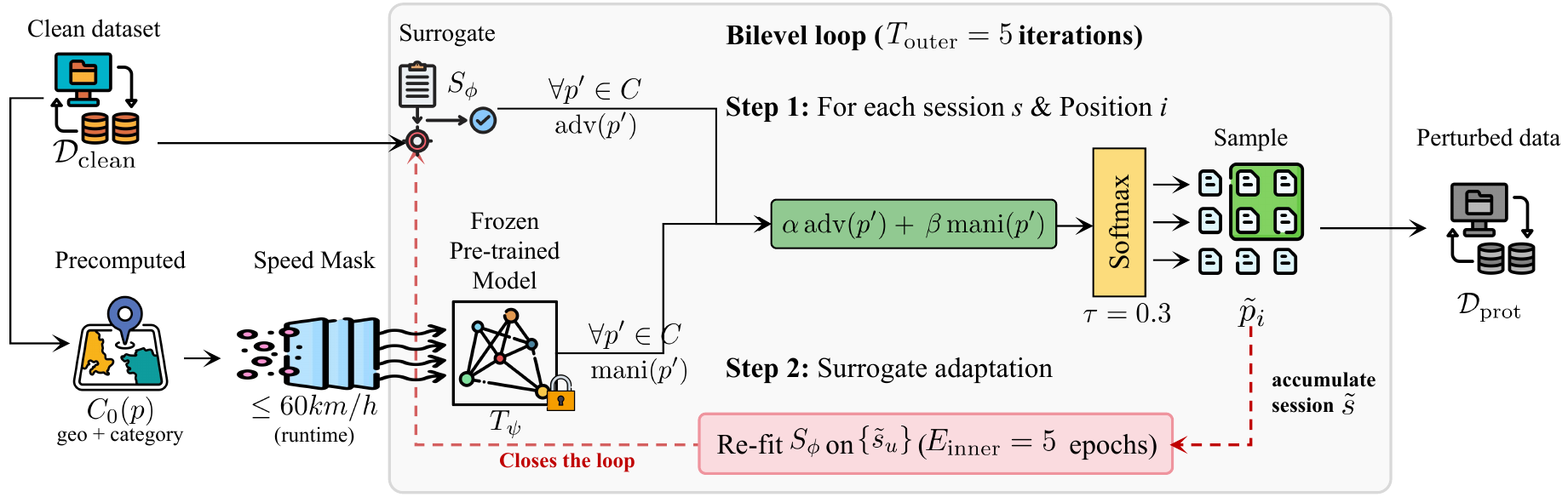}
    \caption{The \ghost framework. \textbf{Inputs}: clean dataset $\mathcal{D}_{\text{clean}}$ initializes the surrogate (warm-up) and seeds the precomputed candidate set $C_0(p)$ (geo + category constraints). \textbf{Runtime}: a speed mask further restricts $C_0$ at each position based on the previously-chosen perturbed POI. \textbf{Bilevel loop}: Step 1 scores every candidate by combining $S_\phi$'s adversarial signal with $T_\psi$'s manifold prior, sampling one $\tilde{p}_i$ per position; Step 2 re-fits $S_\phi$ on the accumulated perturbed sessions. After $T_{\text{outer}}$ rounds the protected dataset $\mathcal{D}_{\text{prot}}$ is released.}
    \label{fig:framework}
\end{figure*}

\section{The \ghost Framework}
\label{sec:method}

\subsection{Overview}
\label{sec:overview}
\ghost\ is a bilevel optimization that produces a perturbed training set $\mathcal{D}_{\text{prot}}$ from the clean dataset $\mathcal{D}_{\text{clean}}$ through position-wise POI substitution (see Figure~\ref{fig:framework}). Each outer round repeats two steps. \textbf{(1) Position-wise scoring} evaluates every candidate substitution at every perturbable position by combining an adversarial signal that measures how much the substitution would damage next-POI supervision with a manifold signal that measures how natural the substitution looks under a frozen real-trajectory likelihood model. \textbf{(2) Surrogate adaptation} re-fits a small victim model on the updated perturbed data so that the next round of scoring reflects a victim that has already absorbed the previous round's perturbation. The two signals are designed to discharge the two failure modes identified: the adversarial signal alone would yield a deterministic error-maximizing map of the kind that purification adversaries are designed to invert, the manifold signal alone would yield benign on-manifold substitutions that protect nothing, and their linear combination produces a map that is simultaneously hard to learn from and hard to invert.

\subsection{Plausibility-Filtered Candidate Set}
\label{sec:candidates}
Hard plausibility constraints define a candidate set per anchor POI before any adversarial optimization begins. For each $p \in \mathcal{P}$ we precompute
\begin{equation}
C_0(p) = \{ p' \in \mathcal{P} :
  d_{\text{hav}}(p, p') \le R,\ \mathrm{cat}(p') = \mathrm{cat}(p),\ p' \neq p \},
\label{eq:candidate}
\end{equation}
truncated to the $K = 32$ POIs closest to $p$ in haversine distance. At $R = 1\,\text{km}$ this yields a per-POI mean candidate-set size of $7.3$ on NYC and $8.7$ on TKY, which is expressive enough for the bilevel scoring to discriminate while small enough to keep adversarial scoring cheap. Sparse-neighbourhood anchors with $|C_0(p)| < 4$ trigger a radius widening, multiplying $R$ by $1.5$ up to four times. The category constraint is never relaxed, even at the cost of leaving a handful of rural anchors with fewer than four candidates, which preserves the $\text{cat\_match} = 1.0$ invariant exactly.

Speed plausibility cannot be precomputed because it depends on the previously-chosen perturbed POI and on the timestamp gap. We enforce it as a context-dependent runtime mask. Given the previously-chosen $\tilde p_{i-1}$ at time $t_{i-1}$ and the current position's timestamp $t_i$, any candidate with implied haversine speed $d_{\text{hav}}(\tilde p_{i-1}, p') / ((t_i - t_{i-1})/3600) > 60\,\text{km/h}$ is excluded from the sampling distribution at position $i$. The runtime-filtered candidate set is denoted $C(p_i; \tilde s_{<i})$, or simply $C(p)$ when the context is unambiguous.

\subsection{Manifold-Prior Substitution Score}
\label{sec:score}
\ghost maintains two transformer networks with identical architecture. The surrogate $S_\phi$ is trained inside the bilevel inner loop on the currently perturbed data and plays the role of the defender's approximation of the deployed victim. The trajectory language model $T_\psi$ is pre-trained on the clean training split and frozen thereafter, encoding the prior over real human check-in sequences. Validation and test sessions are excluded from $T_\psi$'s training to prevent naturalness-signal leakage across splits. 

For each position $i$ in session $s$, with current perturbed history $\tilde{s}_{\le i}$ and candidate $p' \in C(p_i; \tilde s_{<i})$, the substitution score is the linear combination
\begin{align}
\text{adv}(p') &= -\log S_\phi(p_{i+1} \mid \tilde{s}_{<i}, p' \text{ at } i), \label{eq:adv} \\
\text{mani}(p') &= \phantom{-}\log T_\psi(p' \mid \tilde{s}_{<i}), \label{eq:mani} \\
\text{score}(p') &= \alpha \cdot \text{adv}(p') + \beta \cdot \text{mani}(p'). \label{eq:score}
\end{align}

\textbf{Adversarial term.}\quad The adversarial term Eq.~\eqref{eq:adv} is the negative log-likelihood the surrogate assigns to the \emph{clean} next-POI label $p_{i+1}$ after candidate $p'$ has been substituted at position $i$. A candidate that makes the genuine next location surprising under $S_\phi$ maximizes the cross-entropy loss that the deployed victim would incur on the (substituted history, clean target) pair, which is exactly the supervised signal the publisher wants to corrupt. The target is the clean next POI rather than the candidate itself because the deployed victim will be evaluated on clean targets, so the publisher must damage prediction of those specific labels. Boundary positions use $p_{i+1}$ as the target when $i = 0$ and $p_{i-1}$ when $i = L-1$, so that every position contributes to the supervision.

\textbf{Manifold term.}\quad The manifold term Eq.~\eqref{eq:mani} is the trajectory LM's log-likelihood of $p'$ as the next POI given the strict perturbed prefix. Substitutions that lie in the high-density region of the real-trajectory distribution are favoured, and substitutions that look like noise are penalized. Computationally, the term is essentially free, since $T_\psi$ is evaluated once per position to produce a full next-POI distribution, and the log-probabilities of all candidates are gathered from that single forward pass. The manifold cost is therefore $\Theta(1)$ per position in $|C|$, in contrast to the adversarial term, which requires $|C|$ separate surrogate forwards.

\textbf{Sampling.}\quad The candidate is drawn from the low-temperature softmax
\begin{equation}
P(p' \mid p_i, \tilde{s}_{<i}) \propto \exp\left(\text{score}(p') / \tau \right),
\quad \tau = 0.3.
\label{eq:sample}
\end{equation}
The sampling temperature is low so that the score function dominates, but stochasticity is retained for purification resistance. No explicit entropy floor or uniform mixing is used. The manifold term itself contributes the dispersion needed to defeat deterministic-map adversaries, since multiple plausible candidates typically receive comparable manifold scores and the resulting softmax distribution is naturally high-entropy in the high-density region. This dispenses with the entropy-floor heuristic common in randomized adaptations of error-minimizing UE~\cite{huang2021unlearnable,fu2022robust}. Under this heuristic, the sampling softmax is mixed with a uniform distribution at strength $\eta$ (in bits) to enforce a per-position entropy lower bound; the discrete-POI variant is evaluated as GHOST-EF in Section~\ref{sec:ablations}.

\textbf{Score weights.}\quad The weights $(\alpha, \beta) \ge 0$ control the trade-off between adversarial strength and manifold alignment. We set $(\alpha, \beta) = (2.0, 0.5)$ as the recommended default, denoted \ghost, and also report the symmetric setting $(1.0, 1.0)$, denoted GHOST-Sym, as a sensitivity reference. The asymmetric default sits closest to PGD's all-adversarial extreme while retaining the manifold term's purification-resistance benefit. Section~\ref{sec:ablations} sweeps the $(\alpha, \beta)$ grid.

\begin{algorithm}[t]
\caption{\ghost training-data perturbation}
\label{alg:maple}
\begin{algorithmic}[1]
\State \textbf{Input:} clean sessions $\mathcal{D}_{\text{clean}}$,
candidate sets $\{C(p)\}$, frozen trajectory LM $T_\psi$,
weights $(\alpha, \beta)$, temperature $\tau$,
$T_{\text{outer}}$, $E_{\text{inner}}$.
\State Initialize surrogate $S_\phi$ randomly.
\State \textbf{Warm-up:} train $S_\phi$ on $\mathcal{D}_{\text{clean}}$ for $E_{\text{inner}}$ epochs.
\For{$t = 1, \dots, T_{\text{outer}}$}
  \For{each session $s$, each perturbable position $i$}
    \For{each candidate $p' \in C(p_i; \tilde s_{<i})$}
      \State $\text{adv}(p')$, $\text{mani}(p')$
    \EndFor
    \State $\tilde{p}_i \sim
           \mathrm{Categorical}(\exp(\text{score}/\tau))$
  \EndFor
  \State Re-fit $S_\phi$ on $\{\tilde{s}_u\}$ for $E_{\text{inner}}$ epochs.
\EndFor
\State \textbf{return} perturbed dataset $\{\tilde{s}_u\}$.
\end{algorithmic}
\end{algorithm}

\subsection{Bilevel Algorithm}
\label{sec:bilevel}
The full pipeline alternates substitution (Eqs.~\eqref{eq:adv}--\eqref{eq:sample}) with surrogate adaptation for $T_{\text{outer}} = 5$ outer iterations, each followed by $E_{\text{inner}} = 5$ inner epochs of surrogate fine-tuning. Algorithm~\ref{alg:maple} summarizes the procedure; Section~\ref{sec:ablations} sweeps $T_{\text{outer}}$. 
The dominant cost per outer iteration is the adversarial scoring, $\Theta(N \cdot |C| \cdot T_{\text{fwd}})$, where $N$ is the total number of positions and $T_{\text{fwd}}$ is the cost of one surrogate forward pass. The manifold scoring adds $\Theta(N \cdot T_{\text{fwd}})$, identical in order to a single surrogate pass and amounting to roughly $1/|C| \approx 11\text{--}14\%$ relative overhead. Surrogate re-fitting is $\Theta(N \cdot E_{\text{inner}} \cdot T_{\text{fwd}})$. \ghost\ is therefore not asymptotically more expensive than an all-adversarial baseline such as PGD or EM.

\begin{table*}[t]
\centering
\begin{threeparttable}
\caption{Main protection matrix on Foursquare-NYC and Foursquare-TKY. All quantities are computed per-seed and reported as mean $\pm$ std over three seeds. Symbols are defined in the Metrics paragraph (Section~\ref{sec:metrics}). \textbf{Bold} marks the per-column best and \underline{underline} marks the second-best. The recommended \ghost is highlighted.}
\label{tab:main}
\scriptsize
\setlength{\tabcolsep}{2pt}
\begin{tabular}{l ccccc cccccc}
\toprule
\multicolumn{12}{c}{\textbf{Foursquare-NYC}} \\
\cmidrule(lr){1-12}
\textbf{Method} & $\bm{\mathrm{acc}^{(0)}}$ & $\bm{\mathrm{acc}^{(1)}}\downarrow$ & $\bm{\mathrm{acc}^{(2)}}\downarrow$ & $\bm{\mathrm{acc}^{(3)}}\downarrow$ & $\bm{\mathrm{acc}^{(4)}}\downarrow$ & $\bm{\Delta_{\text{prot}}}\uparrow$ & $\bm{\Delta_{\text{surv}}^{(2)}}\downarrow$ & $\bm{\Delta_{\text{surv}}^{(3)}}\downarrow$ & $\bm{\Delta_{\text{surv}}^{(4)}}\downarrow$ & $\bm{\Delta_{\text{mean}}}\uparrow$ & $\bm{\Delta_{\text{worst}}}\uparrow$ \\
\midrule
\rowcolor{blue!8}
\ghost  & 0.1379{\tiny$\pm$0.0009} & \textbf{0.0544}{\tiny$\pm$0.0031} & \underline{0.0672}{\tiny$\pm$0.0038} & \textbf{0.0716}{\tiny$\pm$0.0007} & \textbf{0.0733}{\tiny$\pm$0.0005} & \textbf{0.0835}{\tiny$\pm$0.0032} & 0.0128{\tiny$\pm$0.0039} & 0.0172{\tiny$\pm$0.0026} & \underline{0.0189}{\tiny$\pm$0.0034} & \textbf{0.0735}{\tiny$\pm$0.0017} & \textbf{0.0663}{\tiny$\pm$0.0006} \\
PGD~\cite{madry2018towards}  & 0.1379{\tiny$\pm$0.0009} & \underline{0.0566}{\tiny$\pm$0.0023} & 0.0711{\tiny$\pm$0.0014} & \underline{0.0755}{\tiny$\pm$0.0030} & \underline{0.0771}{\tiny$\pm$0.0020} & 0.0812{\tiny$\pm$0.0022} & 0.0141{\tiny$\pm$0.0016} & \underline{0.0161}{\tiny$\pm$0.0032} & 0.0201{\tiny$\pm$0.0033} & \underline{0.0706}{\tiny$\pm$0.0012} & \underline{0.0645}{\tiny$\pm$0.0004} \\
EM~\cite{huang2021unlearnable}  & 0.1379{\tiny$\pm$0.0009} & 0.0644{\tiny$\pm$0.0014} & 0.0746{\tiny$\pm$0.0043} & 0.0755{\tiny$\pm$0.0006} & 0.0775{\tiny$\pm$0.0013} & 0.0735{\tiny$\pm$0.0006} & \textbf{0.0084}{\tiny$\pm$0.0043} & \textbf{0.0105}{\tiny$\pm$0.0004} & \textbf{0.0130}{\tiny$\pm$0.0003} & 0.0669{\tiny$\pm$0.0017} & 0.0621{\tiny$\pm$0.0008} \\
TS-UE~\cite{jiang2024unlearnable}  & 0.1379{\tiny$\pm$0.0009} & 0.0575{\tiny$\pm$0.0020} & \textbf{0.0668}{\tiny$\pm$0.0043} & 0.0838{\tiny$\pm$0.0032} & 0.0836{\tiny$\pm$0.0020} & \underline{0.0814}{\tiny$\pm$0.0028} & \underline{0.0093}{\tiny$\pm$0.0045} & 0.0263{\tiny$\pm$0.0036} & 0.0261{\tiny$\pm$0.0028} & 0.0695{\tiny$\pm$0.0019} & 0.0551{\tiny$\pm$0.0053} \\
\midrule
\multicolumn{12}{c}{\textbf{Foursquare-TKY}} \\
\cmidrule(lr){1-12}
\textbf{Method} & $\bm{\mathrm{acc}^{(0)}}$ & $\bm{\mathrm{acc}^{(1)}}\downarrow$ & $\bm{\mathrm{acc}^{(2)}}\downarrow$ & $\bm{\mathrm{acc}^{(3)}}\downarrow$ & $\bm{\mathrm{acc}^{(4)}}\downarrow$ & $\bm{\Delta_{\text{prot}}}\uparrow$ & $\bm{\Delta_{\text{surv}}^{(2)}}\downarrow$ & $\bm{\Delta_{\text{surv}}^{(3)}}\downarrow$ & $\bm{\Delta_{\text{surv}}^{(4)}}\downarrow$ & $\bm{\Delta_{\text{mean}}}\uparrow$ & $\bm{\Delta_{\text{worst}}}\uparrow$ \\
\midrule
\rowcolor{blue!8}
\ghost  & 0.1830{\tiny$\pm$0.0007} & \underline{0.0510}{\tiny$\pm$0.0028} & \textbf{0.0899}{\tiny$\pm$0.0017} & \underline{0.0934}{\tiny$\pm$0.0027} & \textbf{0.1014}{\tiny$\pm$0.0042} & \underline{0.1319}{\tiny$\pm$0.0022} & 0.0389{\tiny$\pm$0.0045} & 0.0424{\tiny$\pm$0.0055} & \underline{0.0505}{\tiny$\pm$0.0065} & \underline{0.1048}{\tiny$\pm$0.0015} & \underline{0.0895}{\tiny$\pm$0.0035} \\
PGD~\cite{madry2018towards}  & 0.1830{\tiny$\pm$0.0007} & \textbf{0.0503}{\tiny$\pm$0.0012} & \underline{0.0904}{\tiny$\pm$0.0000}\tnote{$\dagger$} & \textbf{0.0918}{\tiny$\pm$0.0015} & \underline{0.1041}{\tiny$\pm$0.0030} & \textbf{0.1333}{\tiny$\pm$0.0011} & 0.0401{\tiny$\pm$0.0012} & \underline{0.0415}{\tiny$\pm$0.0027} & 0.0539{\tiny$\pm$0.0040} & \textbf{0.1061}{\tiny$\pm$0.0008} & \textbf{0.0917}{\tiny$\pm$0.0017} \\
EM~\cite{huang2021unlearnable}  & 0.1830{\tiny$\pm$0.0007} & 0.0712{\tiny$\pm$0.0028} & 0.0905{\tiny$\pm$0.0011} & 0.0973{\tiny$\pm$0.0002} & 0.1112{\tiny$\pm$0.0037} & 0.1116{\tiny$\pm$0.0016} & \textbf{0.0193}{\tiny$\pm$0.0020} & \textbf{0.0261}{\tiny$\pm$0.0029} & \textbf{0.0400}{\tiny$\pm$0.0051} & 0.0964{\tiny$\pm$0.0004} & 0.0855{\tiny$\pm$0.0016} \\
TS-UE~\cite{jiang2024unlearnable}  & 0.1830{\tiny$\pm$0.0007} & 0.0702{\tiny$\pm$0.0015} & 0.0924{\tiny$\pm$0.0015} & 0.1182{\tiny$\pm$0.0049} & 0.1240{\tiny$\pm$0.0026} & 0.1135{\tiny$\pm$0.0005} & \underline{0.0223}{\tiny$\pm$0.0010} & 0.0480{\tiny$\pm$0.0037} & 0.0538{\tiny$\pm$0.0018} & 0.0901{\tiny$\pm$0.0014} & 0.0655{\tiny$\pm$0.0041} \\
\bottomrule
\end{tabular}
\begin{tablenotes}
\footnotesize
\item[$\dagger$] A zero std reflects seed-stable behaviour rounded to four decimals, not missing data.
\end{tablenotes}
\end{threeparttable}
\end{table*}

\section{Experiments}
\label{sec:exp}

\subsection{Experimental Setup}

\textbf{Datasets.}\quad We evaluate on Foursquare-NYC and Foursquare-TKY from the TSMC2014 release~\cite{yang2015modeling}, the two standard category-aware POI benchmarks~\cite{luo2021stan,yang2022getnext}. After filtering users with fewer than 10 check-ins and POIs visited fewer than 10 times, we segment per-user sessions by 24-hour gaps and hold out $10\%$ of sessions for validation and $20\%$ for test. The resulting NYC split contains 1{,}083 users, 5{,}135 POIs, and 10{,}958 training, 937 validation, and 2{,}359 test sessions. TKY contains 2{,}293 users, 7{,}873 POIs, and 34{,}902 training, 3{,}442 validation, and 6{,}393 test sessions.

\textbf{Settings.}\quad The victim is a two-layer causal Transformer with embedding dimension $d = 128$, four attention heads, and maximum sequence length $L_{\max} = 100$. We train with AdamW (learning rate $10^{-3}$, batch size 256) and early stopping on validation acc@1, retraining a fresh victim from scratch for every combination. The surrogate $S_\phi$ and trajectory LM $T_\psi$ share the same architecture, with $T_\psi$ pre-trained on the clean training split and cached across seeds and ablations. The bilevel outer loop uses $T_{\text{outer}} = 5$ iterations, score weights $(\alpha, \beta) = (2.0, 0.5)$, and sampling temperature $\tau = 0.3$, with the surrogate incrementally fine-tuned on the updated perturbed data at the end of each round. We evaluate the no-adversary baseline (A1) alongside three purification adversaries: the BridgePure denoising-bridge proxy (A2), the frequency-table inverter (A3), and the bigram-adaptive purifier (A4). All purification adversaries operate at the default leak ratio $r = 0.05$. The A2 denoiser is a two-layer causal Transformer trained on leaked pairs with masked cross-entropy and applied with three iterative argmax refinement steps. Main-table experiments report results over three random seeds.

\textbf{Baselines.}\quad We compare against three published methods spanning the main design families of unlearnable examples, all sharing \ghost's candidate set, surrogate architecture, and bilevel budget. \emph{EM}~\cite{huang2021unlearnable} is the canonical error-minimizing UE, adapted to discrete POIs. \emph{TS-UE}~\cite{jiang2024unlearnable} is the closest published sequential-UE method. \emph{PGD}~\cite{madry2018towards} is the deterministic error-maximizing counterpart of \ghost's stochastic score. The internal variants GHOST-Sym ($\alpha{=}\beta{=}1$) and the entropy-floor predecessor GHOST-EF are evaluated only in the manifold-prior ablation.

\textbf{Metrics.}\label{sec:metrics}\quad The victim's predictive quality is measured by top-$K$ accuracy (acc@$K$) and Mean Reciprocal Rank (MRR), following standard next-POI evaluation~\cite{luo2021stan,yang2022getnext}. Under the standard UE protocol~\cite{huang2021unlearnable,wang2024bridgepure}, $\mathrm{acc}^{(0)}$ denotes the victim acc@1 after training on the clean dataset $\mathcal{D}_{\text{clean}}$, $\mathrm{acc}^{(1)}$ on the protected dataset $\mathcal{D}_{\text{prot}}$ (A1, no adversary), and $\mathrm{acc}^{(a)}$ for $a \in \{2, 3, 4\}$ on the dataset recovered by adversary A$a$, where A2 is the BridgePure denoising-bridge proxy, A3 is the frequency-table inverter, and A4 is the bigram-adaptive purifier. The aggregated $\Delta_{\text{mean}}$ and $\Delta_{\text{worst}}$ below pool A1--A3, with A4 reported separately as an adaptive sensitivity check. Lower $\mathrm{acc}^{(a)}$ indicates stronger protection. We define four protection scores:
\begin{align}
\Delta_{\text{prot}}       &= \mathrm{acc}^{(0)} - \mathrm{acc}^{(1)},                           \label{eq:prot-gap} \\
\Delta_{\text{surv}}^{(a)} &= \mathrm{acc}^{(a)} - \mathrm{acc}^{(1)},                           \label{eq:surv-gap} \\
\Delta_{\text{mean}}       &= \mathrm{acc}^{(0)} - \tfrac{1}{3}\sum_{a=1}^{3}\mathrm{acc}^{(a)}, \label{eq:mean-eff} \\
\Delta_{\text{worst}}      &= \mathrm{acc}^{(0)} - \max_{a \in \{1,2,3\}} \mathrm{acc}^{(a)},    \label{eq:worst-eff}
\end{align}
referred to respectively as the \emph{protection gap}, the \emph{survival gap} under adversary $a$, and the mean and worst-case attacker-aggregated protection scores. Higher $\Delta_{\bullet}$ indicates stronger protection. We treat $\Delta_{\text{mean}}$ as the headline since $\Delta_{\text{worst}}$ favours methods whose protected accuracy is already high. We additionally report the category-match rate, geographic-violation rate, mean and 95th-percentile haversine displacements, and substitution rate, all standard in trajectory-substitution privacy~\cite{abul2008never,he2015dpt}. The relative ranking of methods is preserved across acc@1, acc@5, and MRR.

\subsection{Protection Results}
\label{sec:main}

The main results matrix for the four methods, both datasets, and the four attacker postures (A1--A4) is presented in Table~\ref{tab:main}. The clean victim achieves acc@1 of $0.1379 \pm 0.0009$ on NYC and $0.1830 \pm 0.0007$ on TKY, matching published numbers for STAN on the same splits and confirming our victim is competitive. We focus on acc@1 throughout; the relative ranking of methods is preserved on acc@5 and MRR.

\paragraph{Headline metric}
The rightmost two columns of Table~\ref{tab:main} report the attacker-aggregated protection scores. We treat $\Delta_{\text{mean}}$ as the primary headline because $\Delta_{\text{worst}}$ is mechanically inflated by methods whose protected accuracy is already high: a defender that ``protects'' by leaving the data nearly unchanged trivially wins $\Delta_{\text{worst}}$ while providing no real protection. On NYC, \ghost delivers the best $\Delta_{\text{mean}}$ ($0.0735$) and $\Delta_{\text{worst}}$ ($0.0663$), leading PGD by $0.29$\,pp and $0.18$\,pp respectively. On TKY, PGD and \ghost are statistically indistinguishable on $\Delta_{\text{mean}}$ (within $1\sigma$), and both clearly dominate EM and TS-UE.

\begin{figure}[t]
\centering
\includegraphics[width=\columnwidth]{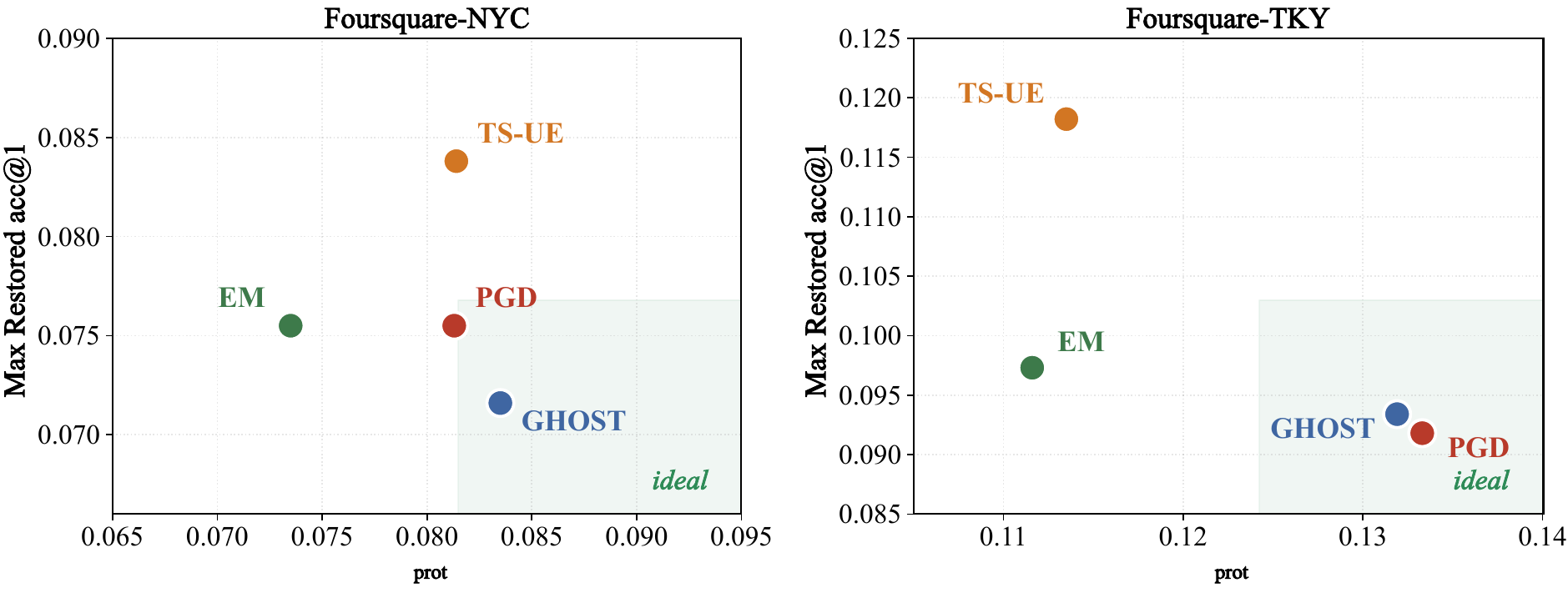}
\caption{Protection-versus-purification-resistance Pareto plot. The horizontal axis is the protection gap $\Delta_{\text{prot}}$ (higher is stronger protection); the vertical axis is the maximum restored acc@1 across attackers A1, A2, A3 (lower is stronger purification resistance). The shaded lower-right corner marks the ideal regime. On Foursquare-NYC, \ghost\ is the unique Pareto-optimal method and sits closest to the ideal corner. On Foursquare-TKY, \ghost\ and PGD jointly occupy the Pareto front within one per-cell standard deviation, and both clearly dominate EM and TS-UE.}
\label{fig:pareto}
\end{figure}

\paragraph{Pareto-optimal trade-off}
The fundamental trade-off for any UE method is between the accuracy a naive learner is denied ($\Delta_{\text{prot}}$) and the maximum accuracy a purification adversary can recover ($\max_a \mathrm{acc}^{(a)}$). Every (method, dataset) configuration is plotted on this plane in Figure~\ref{fig:pareto}. The lower-right corner is ideal, with high gap and low restored. On Foursquare-NYC, \ghost dominates PGD on both axes, with higher $\Delta_{\text{prot}}$ ($0.0835$ vs $0.0812$) and lower max-restored accuracy ($0.0716$ vs $0.0755$). On Foursquare-TKY, \ghost and PGD jointly occupy the Pareto front within one per-cell standard deviation. Methods strictly dominated by \ghost on at least one dataset include EM and TS-UE.

\begin{figure*}[t]
\centering
\includegraphics[width=1.0\linewidth]{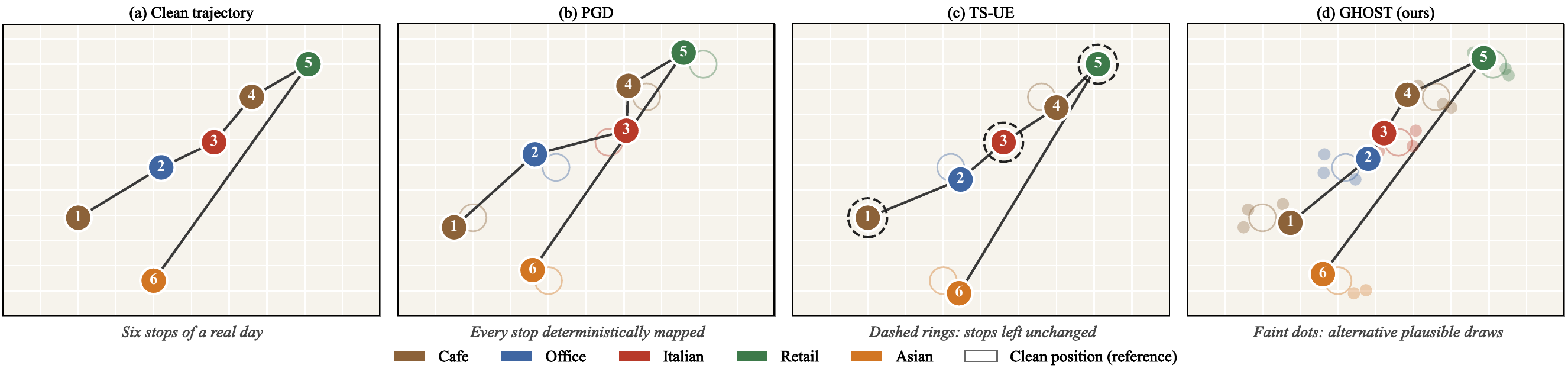}
\caption{Illustrative qualitative comparison on a Manhattan day trajectory. Filled markers show each method's released POI; faint hollow rings mark the original clean POI for reference; colour encodes category. \textbf{(a)} A six-stop clean session. \textbf{(b)} PGD's deterministic argmin pulls every position toward the same surrogate-hardest alternative, producing a fixed and therefore invertible map. \textbf{(c)} TS-UE's embedding-space PGD followed by snap-to-candidate leaves about a quarter of positions unchanged (perturb rate $0.749$ on NYC), yielding weak protection. \textbf{(d)} \ghost\ samples each substitution from a manifold-weighted softmax over plausible candidates, so the released trajectory remains visually coherent with the clean one while every position is genuinely perturbed. The example uses real Manhattan landmark coordinates and serves as a qualitative aid.}
\label{fig:qualitative}
\end{figure*}

\begin{table}[t]
\centering
\caption{Plausibility metrics. Geo. Viol.\ is the fraction of substitutions exceeding the 1\,km nominal radius after sparse-area widening; P95 denotes the 95th percentile of the per-substitution haversine displacement; mean and P95 displacements are in kilometres.}
\label{tab:plausibility}
\setlength{\tabcolsep}{3pt}
\footnotesize
\begin{tabular}{l l c c c c}
\toprule
\textbf{Dataset} & \textbf{Method} & \textbf{Sub. Rate} & \textbf{Geo. Viol.} & \textbf{Mean Disp.} & \textbf{P95 Disp.} \\
\midrule
\multirow{4}{*}{NYC}
& \cellcolor{blue!8}\ghost & \cellcolor{blue!8}0.861 & \cellcolor{blue!8}0.6339{\tiny$\pm$0.0016} & \cellcolor{blue!8}1.898{\tiny$\pm$0.006} & \cellcolor{blue!8}4.714{\tiny$\pm$0.007} \\
& PGD                     & 0.861 & 0.6464{\tiny$\pm$0.0002} & 1.934{\tiny$\pm$0.015} & 4.723{\tiny$\pm$0.018} \\
& EM                      & 0.861 & 0.6130{\tiny$\pm$0.0016} & 1.845{\tiny$\pm$0.005} & 4.684{\tiny$\pm$0.003} \\
& TS-UE                   & 0.749 & 0.6321{\tiny$\pm$0.0060} & 1.909{\tiny$\pm$0.021} & 4.702{\tiny$\pm$0.017} \\
\midrule
\multirow{4}{*}{TKY}
& \cellcolor{blue!8}\ghost & \cellcolor{blue!8}0.969 & \cellcolor{blue!8}0.4128{\tiny$\pm$0.0023} & \cellcolor{blue!8}1.241{\tiny$\pm$0.003} & \cellcolor{blue!8}4.172{\tiny$\pm$0.037} \\
& PGD                     & 0.969 & 0.4434{\tiny$\pm$0.0015} & 1.355{\tiny$\pm$0.002} & 4.307{\tiny$\pm$0.017} \\
& EM                      & 0.969 & 0.3936{\tiny$\pm$0.0018} & 1.187{\tiny$\pm$0.008} & 4.101{\tiny$\pm$0.019} \\
& TS-UE                   & 0.855 & 0.4147{\tiny$\pm$0.0099} & 1.245{\tiny$\pm$0.024} & 4.169{\tiny$\pm$0.051} \\
\bottomrule
\end{tabular}
\end{table}

\paragraph{Adaptive sequence-aware adversary}
The fourth attacker posture (A4) reported in Table~\ref{tab:main} is a
\emph{sequence-aware adaptive adversary} that exploits one step of
perturbed context. From the same leaked pairs as A2 and A3 the attacker
builds a bigram conditional table
$\hat p(p^{\text{clean}}_i \mid p^{\text{prot}}_i, p^{\text{prot}}_{i-1})$
and falls back to the unigram table when the bigram key is unseen.
\ghost yields the lowest A4-restored accuracy on both
datasets ($0.0733$ on NYC and $0.1014$ on TKY, lower than PGD by
$0.4$\,pp and $0.3$\,pp respectively and lower than TS-UE by $1.0$\,pp
and $2.3$\,pp).

\subsection{Per-Method Failure Modes}
\label{sec:attackers}

Three baseline failure modes that motivate \ghost's design are visible in Table~\ref{tab:main}.

\paragraph{TS-UE} Strong on BridgePure and catastrophic under
A3. TS-UE's error-min PGD in surrogate embedding space, followed by snap-to-candidate, produces perturbations that approximate the manifold direction well enough to make BridgePure-restored acc@1 the lowest on NYC ($0.0668$). The perturbation is also \emph{deterministic} at the position level, so the frequency-table A3 inverter recovers $0.0838$ on NYC and $0.1182$ on TKY, by far the worst A3 numbers in Table~\ref{tab:main}. $\Delta_{\text{worst}}$ is therefore dominated by A3 for TS-UE, dragging it to last place ($0.0551$ on NYC, $0.0655$ on TKY).

\paragraph{PGD} Aggressive protection with easier purification. 
PGD's all-adversarial argmin yields the lowest protected acc@1 on both datasets ($0.0566$ on NYC, $0.0503$ on TKY). The determinism of PGD makes both adversaries partially effective. On NYC, the BridgePure-restored accuracy climbs to $0.0711$ ($\Delta_{\text{surv}}^{(2)} = 0.0141$), and the frequency-table-restored accuracy climbs to $0.0755$ ($\Delta_{\text{surv}}^{(3)} = 0.0161$). PGD's $\Delta_{\text{mean}}$ is comparable to \ghost on TKY but $0.29$\,pp below on NYC.

\paragraph{EM} Naturalistic substitutions with the weakest protection. EM's error-minimizing objective rewards on-manifold candidates with high victim-predicted likelihood, yielding the most plausible-looking perturbations but the weakest protection. EM has the highest protected acc@1 on both datasets ($0.0644$ NYC, $0.0712$ TKY) and therefore the smallest protection gap ($\Delta_{\text{prot}} = 0.0735$ on NYC, $0.1116$ on TKY). Its low survival gaps are a mechanical consequence of weak protection rather than evidence of purification resistance, since when little has been protected, little remains to be restored.

\paragraph{\ghost} Combines high protection-gap with low restored accuracy. The failure modes above demonstrate that no single design choice achieves both strong protection and purification resistance at the same time. \ghost's manifold-aligned stochastic substitutions break this trade-off. \ghost attains the lowest protected acc@1 on NYC, the second-lowest on TKY, and restored accuracy within $0.5$\,pp of the lowest in every (dataset, attacker) cell. The one baseline that occasionally undercuts \ghost on the restored axis (TS-UE on NYC-A2) pays for it elsewhere by collapsing on A3 with a restored acc@1 of $0.0838$, the worst in the table. \ghost is therefore the only method that delivers a top-tier protection-gap together with the lowest A4 (bigram adaptive) restored accuracy on both datasets.

\subsection{Plausibility Comparison}
\label{sec:plausibility}

A protection that fails plausibility is unusable. Downstream services break, and external auditors detect the manipulation. Each method's per-position substitution behavior is illustrated qualitatively on a real Manhattan trajectory (Figure~\ref{fig:qualitative}) and quantified across the full benchmark (Table~\ref{tab:plausibility}).

\textbf{Per-position plausibility.} All four methods achieve a category-match rate of $1.000 \pm 0.000$ on both datasets, which confirms that the shared candidate-set construction enforces the semantic plausibility invariant absolutely. The remaining metrics quantify how aggressively each method moves POIs within the plausible candidate set. The geographic-violation rate and the mean haversine displacement lie in a tight band across methods. On NYC, the ranges are $0.615$--$0.646$ and $1.85$--$1.94$\,km; on TKY, $0.394$--$0.443$ and $1.19$--$1.35$\,km. Three of the four methods substitute at the candidate-set's natural substitution rate of $0.861$ on NYC and $0.969$ on TKY. The exception is TS-UE, whose embedding-space PGD followed by snap-to-candidate substitutes at a rate of $0.749$ on NYC and $0.855$ on TKY, roughly $11$ percentage points below the other methods because the $L_2$-nearest candidate after PGD often coincides with the original POI. \ghost\ sits squarely on the plausibility Pareto front with PGD and the other baselines. The algorithmic contribution is concentrated in which candidate is \emph{chosen} from the plausible neighbourhood, not in the construction of the neighbourhood itself.

\textbf{Sequence-level naturalness.} The plausibility metrics above measure single-position properties. A complementary \emph{sequence-level} naturalness score, the mean per-position log-likelihood of the perturbed sequence under the frozen trajectory language model $T_\psi$, is reported in Table~\ref{tab:naturalness}. Clean trajectories anchor the upper bound ($-3.83$ on NYC, $-3.93$ on TKY). On NYC \ghost\ attains the highest naturalness among all unlearnable-examples methods ($-10.02$), a $3.5\%$ improvement over PGD. On TKY, the error-minimising baselines TS-UE and EM achieve higher naturalness than \ghost\ (their score function actively rewards high-LM-likelihood substitutions, which trades adversarial strength for naturalness), while \ghost\ remains substantially more natural than PGD. The pattern matches the protection-vs-restored Pareto picture in Figure~\ref{fig:pareto}, where the methods that beat \ghost\ on naturalness lose to it on protection-gap and on BridgePure-restored accuracy.

\begin{table}[t]
\centering
\caption{Mean per-position $\log p_{T_\psi}$ under the frozen trajectory LM. Higher values are closer to natural human mobility, and clean trajectories provide the upper-bound reference.}
\label{tab:naturalness}
\setlength{\tabcolsep}{6pt}
\footnotesize
\begin{tabular}{l c c}
\toprule
\textbf{Method} & \textbf{Foursquare-NYC}$\uparrow$ & \textbf{Foursquare-TKY}$\uparrow$ \\
\midrule
\textit{Clean (reference)} & -3.8272 & -3.9270 \\
\midrule
\cellcolor{blue!8}\ghost  & \cellcolor{blue!8}\textbf{-10.0208}{\tiny$\pm$0.0132} & \cellcolor{blue!8}-9.2804{\tiny$\pm$0.0586} \\
TS-UE                     & \underline{-10.0666}{\tiny$\pm$0.0189} & \textbf{-8.5242}{\tiny$\pm$0.0423} \\
EM                        & -10.1250{\tiny$\pm$0.0363} & \underline{-8.6316}{\tiny$\pm$0.0593} \\
PGD                       & -10.3664{\tiny$\pm$0.0261} & -9.7827{\tiny$\pm$0.0803} \\
\bottomrule
\end{tabular}
\end{table}

\subsection{Ablation Study}
\label{sec:ablations}

\paragraph{Manifold prior is the active ingredient}
The role of the manifold prior is isolated in Table~\ref{tab:manifold-ablation} by comparing \ghost\ against GHOST-Sym and GHOST-EF.Both manifold-prior variants outperform the entropy-floor variant on protected acc@1 by $0.5$ to $1.2$\,pp, confirming that the manifold prior is the substantive contribution. \ghost\ additionally beats GHOST-Sym by $0.75$\,pp on protected acc@1 and by $0.21$\,pp on the BridgePure-restored axis, motivating the recommended $(\alpha, \beta) = (2.0, 0.5)$ default.

\begin{table}[t]
\centering
\caption{Manifold-prior ablation on Foursquare-NYC. GHOST ($\alpha{=}2.0$, $\beta{=}0.5$) is compared against GHOST-Sym (symmetric weights $\alpha{=}\beta{=}1$) and against GHOST-EF, a baseline variant that replaces the manifold prior with a 3-bit entropy floor on the sampling softmax.}
\label{tab:manifold-ablation}
\setlength{\tabcolsep}{2pt}
\footnotesize
\begin{tabular}{l c c c c c}
\toprule
\textbf{Variant} & \textbf{Manifold} & \textbf{Entropy Floor} & $\bm{\mathrm{acc}^{(1)}}\downarrow$ & $\bm{\mathrm{acc}^{(2)}}\downarrow$ & $\bm{\mathrm{acc}^{(3)}}\downarrow$ \\
\midrule
\cellcolor{blue!8}\ghost  & \cellcolor{blue!8}\checkmark & \cellcolor{blue!8}$\times$ & \cellcolor{blue!8}\textbf{0.0544} & \cellcolor{blue!8}\textbf{0.0672} & \cellcolor{blue!8}\textbf{0.0716} \\
GHOST-Sym                 & \checkmark & $\times$ & \underline{0.0619} & \underline{0.0693} & \underline{0.0715} \\
GHOST-EF                  & $\times$   & 3 bits   & 0.0666 & 0.0762 & 0.0712 \\
\bottomrule
\end{tabular}
\end{table}

\paragraph{Score weighting $(\alpha, \beta)$}
Four corners of the $(\alpha, \beta)$ grid are evaluated in Table~\ref{tab:ab}. Increasing $\alpha$ from $1.0$ to $2.0$, with $\beta$ correspondingly reduced, lifts $\Delta_{\text{prot}}$ from $0.0781$ at the symmetric default to $0.0810$ at $(2.0, 0.5)$. Increasing $\beta$ above $\alpha$ (that is, $(1.0, 1.5)$) instead \emph{degrades} protection by $0.52$\,pp and inflates $\Delta_{\text{surv}}^{(2)}$ from $0.0091$ to $0.0197$, consistent with the analysis in Section~\ref{sec:analysis}. Too much manifold weight produces benign substitutions whose surrogate-hardness is sub-optimal, and the denoiser's job becomes easier because it can simply identify the manifold mode the substitution is shifted toward. The recommended $(2.0, 0.5)$ corner wins three of four columns, including the headline $\Delta_{\text{mean}}$.

\begin{table}[t]
\centering
\caption{Score-weight sensitivity on Foursquare-NYC. \textbf{Bold} marks the per-column best, and \underline{underline} the second-best.}
\label{tab:ab}
\setlength{\tabcolsep}{6pt}
\footnotesize
\begin{tabular}{l c c c c}
\toprule
$\bm{(\alpha, \beta)}$ & $\bm{\Delta_{\text{prot}}}\uparrow$ & $\bm{\Delta_{\text{surv}}^{(2)}}\downarrow$ & $\bm{\Delta_{\text{surv}}^{(3)}}\downarrow$ & $\bm{\Delta_{\text{mean}}}\uparrow$ \\
\midrule
(1.0, 1.0)        & 0.0781              & \underline{0.0091} & 0.0117              & \underline{0.0712} \\
(1.0, 1.5)        & 0.0729              & 0.0197              & \underline{0.0113} & 0.0626 \\
(1.5, 1.0)        & \underline{0.0786}  & 0.0144              & \textbf{0.0106}     & 0.0703 \\
\rowcolor{blue!8}
\textbf{(2.0, 0.5)} & \textbf{0.0810}   & \textbf{0.0076}     & 0.0146              & \textbf{0.0736} \\
\bottomrule
\end{tabular}
\end{table}

\paragraph{Entropy floor}
A central claim is that the manifold term \emph{replaces} the entropy-floor randomization used in prior UE work~\cite{huang2021unlearnable,fu2022robust}. We test this prediction by sweeping the entropy-floor strength $\eta$ with $(\alpha, \beta) = (2.0, 0.5)$ held fixed, with results in Table~\ref{tab:entropy}. Setting $\eta = 0$ attains the best $\Delta_{\text{prot}}$, $\Delta_{\text{mean}}$, and $\Delta_{\text{worst}}$ simultaneously, and the three aggregated scores all trend downward as $\eta$ increases. The manifold term is sufficient, and the legacy randomization is redundant and actively harmful.

\begin{table}[t]
\centering
\caption{Entropy-floor ablation on Foursquare-NYC ($(\alpha, \beta) = (2.0, 0.5)$ fixed). \textbf{Bold} marks the per-column best, and \underline{underline} the second-best.}
\label{tab:entropy}
\setlength{\tabcolsep}{6pt}
\footnotesize
\begin{tabular}{l c c c}
\toprule
$\bm{\eta}$ \textbf{(bits)} & $\bm{\Delta_{\text{prot}}}\uparrow$ & $\bm{\Delta_{\text{mean}}}\uparrow$ & $\bm{\Delta_{\text{worst}}}\uparrow$ \\
\midrule
\rowcolor{blue!8}
\textbf{0} & \textbf{0.0837}    & \textbf{0.0763}    & \textbf{0.0691} \\
1          & \underline{0.0777} & \underline{0.0713} & \underline{0.0680} \\
2          & 0.0731              & 0.0685              & 0.0641 \\
3          & 0.0718              & 0.0688              & 0.0651 \\
\bottomrule
\end{tabular}
\end{table}

\begin{figure}[t]
\centering
\includegraphics[width=\columnwidth]{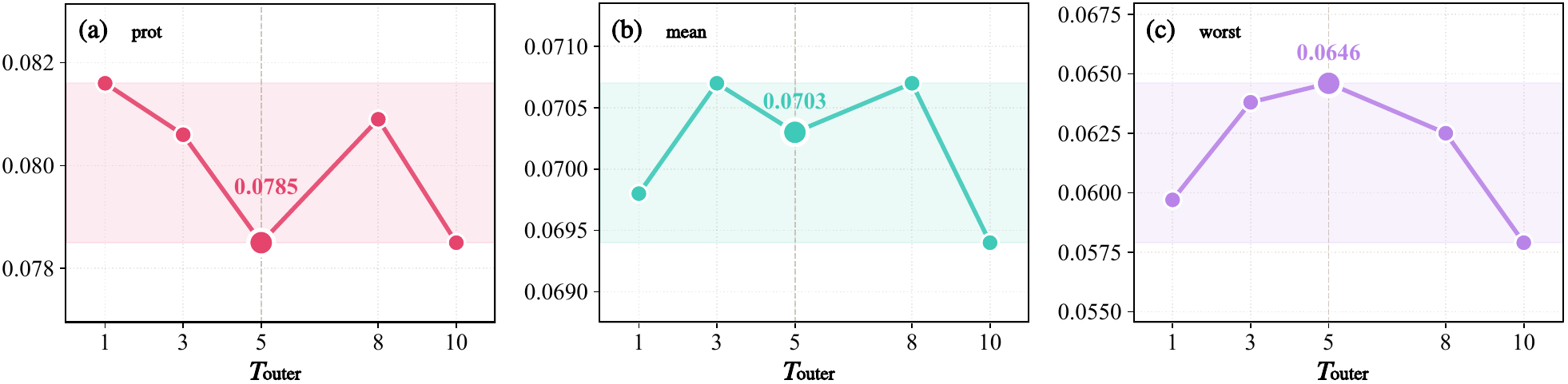}
\caption{Bilevel convergence on Foursquare-NYC ($(\alpha, \beta) = (2.0, 0.5)$). The three aggregated protection scores ($\Delta_{\text{prot}}$, $\Delta_{\text{mean}}$, $\Delta_{\text{worst}}$) cluster within narrow bands across the sweep, confirming that the bilevel procedure converges after a small number of outer rounds. The default $T_{\mathrm{outer}} = 5$ (dashed line) attains the highest $\Delta_{\mathrm{worst}}$, motivating it as the recommended setting.}
\label{fig:tconv}
\end{figure}

\paragraph{Bilevel convergence}
The number of bilevel outer iterations $T_{\text{outer}} \in \{1, 3, 5, 8, 10\}$ is swept with $(\alpha, \beta) = (2.0, 0.5)$ held fixed (see Figure~\ref{fig:tconv}). All five $\Delta_{\text{mean}}$ values lie within $0.0013$ of one another and the three aggregated scores show no monotonic trend, confirming that the bilevel procedure converges after a small number of outer rounds. We adopt $T_{\text{outer}} = 5$ as the default. Single-round ($T_{\text{outer}} = 1$) yields the largest $\Delta_{\text{prot}}$ but the smallest $\Delta_{\text{worst}}$, since the surrogate has not yet absorbed the perturbation and the resulting map is easier for a purification adversary to invert in the worst case.

\begin{figure}[t]
\centering
\includegraphics[width=1.0\linewidth]{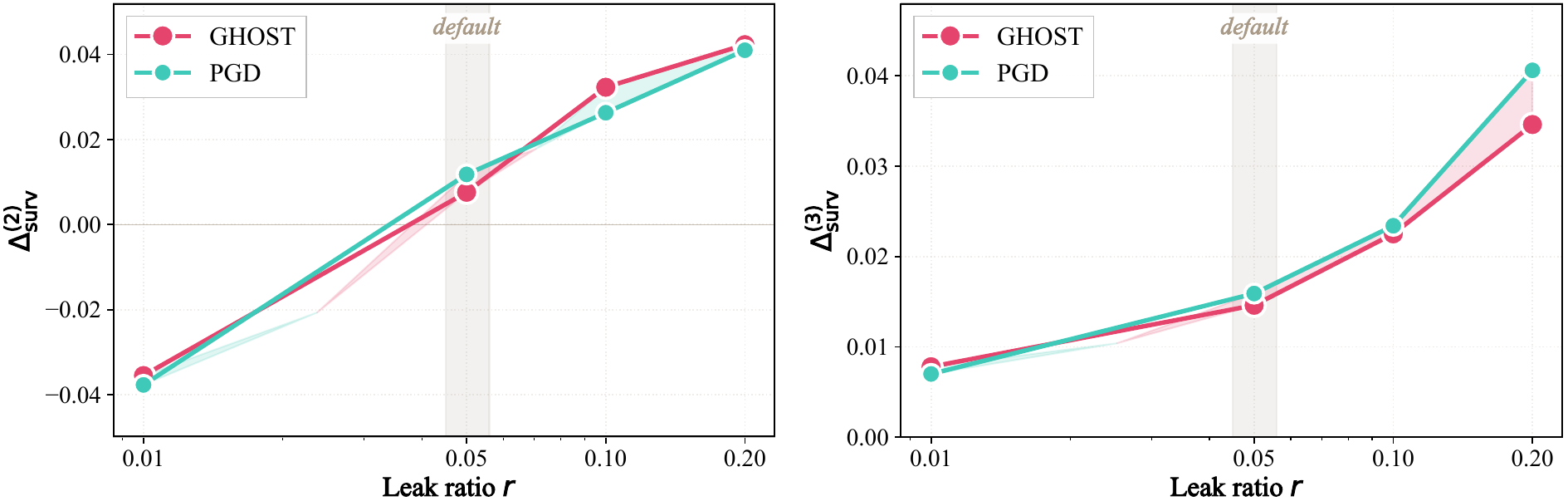}
\caption{Leak-ratio sensitivity on Foursquare-NYC ($(\alpha, \beta) = (2.0, 0.5)$ fixed). \textbf{Left:} survival gap under the BridgePure-proxy adversary (A2); the two curves cross near $r = 0.10$, with \ghost\ winning at the default $r = 0.05$ and PGD slightly winning at higher leak. \textbf{Right:} survival gap under the frequency-table adversary (A3); \ghost\ stays below PGD across the entire sweep, with the gap widening at $r = 0.20$. Lower is better. Shading between the curves indicates the winning method at each leak ratio (blue for \ghost, red for PGD). Negative values at $r = 0.01$ mean the adversary's restoration is worse than no restoration.}
\label{fig:leak}
\end{figure}

\paragraph{Leak ratio sensitivity}
The attacker's leak ratio $r$ is swept from $1\%$ to $20\%$ on NYC (see Figure~\ref{fig:leak}). At $r = 0.01$, both \ghost and PGD have \emph{negative} BridgePure-proxy survival gaps. The denoiser, trained on so few pairs, actively damages the data. At $r = 0.05$, the default in our main tables, \ghost matches or beats PGD on both adversaries. At $r = 0.10$, \ghost's A2 survival gap rises slightly above PGD's ($0.0323$ vs.\ $0.0263$), reflecting that with more leaked data the denoiser begins to imitate the \ghost substitution distribution. \ghost still wins on A3 ($0.0225$ vs.\ $0.0234$). At $r = 0.20$, both methods degrade significantly, but \ghost retains a consistent edge on A3 ($0.0346$ vs.\ $0.0406$). Across the full sweep, \ghost's A3 survival gap is always lower than PGD's, confirming the analysis in Section~\ref{sec:analysis}. The non-bijective on-manifold map is uniformly harder to invert by a context-free frequency table, regardless of leakage.

\subsection{Computational Cost}
The dominant cost of \ghost is the adversarial candidate scoring, which requires one surrogate forward pass per candidate per position per outer iteration. With $|C| \approx 7\text{--}9$ and $T_{\text{outer}} = 5$, each outer round costs $\Theta(N|C| T_{\text{fwd}})$. This is identical in order to PGD and EM, which also evaluate the surrogate at every candidate. The additional cost specific to \ghost is one trajectory-LM forward per position per outer iteration, because $T_\psi$ produces the full next-POI distribution in a single forward, we gather all candidate log-probabilities at once. The total relative overhead of the manifold term over an all-adversarial baseline is therefore $1/|C| \approx 11\text{--}14\%$. The trajectory LM itself is pre-trained once per dataset on the clean training split and disk-cached for reuse across seeds and ablations, so the amortized cost is negligible across our full experiment matrix. We confirm empirically that the entire main matrix (four methods $\times$ four attackers $\times$ three seeds $\times$ two datasets, plus several ablation variants on NYC) completes in under one day of wall-clock on a single NVIDIA A100. Reproducing all ablations adds approximately one further day.

\subsection{Why Manifold-Prior Defeats Purification}
\label{sec:analysis}
The empirical pattern in Table~\ref{tab:main} and Figure~\ref{fig:pareto} admits a structural explanation that ties the design of \ghost back to the threat model of Section~\ref{sec:threat}. We make the argument informally. It is the design intuition that motivates the framework, supported here by the numbers just reported.

\textbf{Optimal denoiser collapses on on-manifold inputs.}\quad A denoising adversary trained on leaked $(p^{\text{clean}}, p^{\text{prot}})$ pairs is, in the population limit, the conditional expectation $\mathbb{E}[p^{\text{clean}} \mid p^{\text{prot}}]$ or its argmax. Its behaviour depends entirely on the joint distribution induced by the protection map. If the map shifts inputs to low-density regions, as additive noise on image UE or category-violating substitutions on POI UE both do, then perturbed inputs are statistically distinguishable from clean inputs and the denoiser has a well-defined job, namely to map each off-manifold input back to its nearest on-manifold mode. \ghost's perturbation map, by construction, never leaves the manifold. Every published $p^{\text{prot}}$ is itself a plausible POI that some other user could genuinely have visited in this context, so $\mathbb{E}[p^{\text{clean}} \mid p^{\text{prot}}]$ approaches the marginal prior over plausible POIs at this position and the optimal denoiser approaches the identity. Table~\ref{tab:main} confirms the prediction. \ghost\ attains the lowest BridgePure-restored accuracy on TKY and is within statistical noise of the lowest on NYC; the one baseline with a marginally lower NYC value (TS-UE, $0.0668$) is dominated on A3.

\textbf{Frequency-table inversion degenerates to uniform.}\quad The A3 inverter is the context-free argmax of the same conditional. Two failure modes apply on \ghost's output. \ghost samples stochastically over a manifold-weighted softmax, so a clean POI $p$ maps to multiple plausible $p'$ with comparable probability, and symmetrically a perturbed $p'$ may be the substituted counterpart of several distinct clean POIs that share its neighbourhood. The empirical $\hat{p}(p^{\text{clean}} \mid p^{\text{prot}})$ is therefore high-entropy and its argmax has high error. Because $p^{\text{prot}}$ is itself a plausible POI, the inverter also cannot distinguish "$p^{\text{prot}}$ is a perturbation of some other POI" from "$p^{\text{prot}}$ was genuinely visited." Table~\ref{tab:main} again matches the prediction. \ghost attains the lowest A3-restored accuracy on Foursquare-NYC and is statistically tied with PGD on Foursquare-TKY. The bigram adaptive variant A4 suffers the same degeneration, with \ghost the lowest on both datasets.

\textbf{Why the adversarial term is still required.}\quad The manifold term alone describes a method that publishes naturalistic but un-protective substitutions, since every position would land in the prior's high-density region but no position would be particularly hard for the victim to predict. The adversarial term repositions \ghost within the high-density region, selecting the on-manifold candidate that maximally damages next-POI supervision. The resulting map is hard to denoise because every output is on-manifold, and hard to learn from because every position's clean target is shifted toward the surrogate-hardest plausible alternative. The entropy-floor ablation in Table~\ref{tab:entropy} provides the final piece of evidence. Adding the legacy randomization knob on top of the manifold term degrades $\Delta_{\text{prot}}$, $\Delta_{\text{mean}}$, and $\Delta_{\text{worst}}$ relative to the manifold-only setting ($\eta = 0$), confirming that the manifold prior already supplies the dispersion the entropy floor was designed to provide.

\section{Discussion}
\label{sec:limits}

\textbf{Limitations.}\quad Our evaluation uses a single STAN-like causal Transformer victim, and cross-architecture transferability is left for future work. Discrete-POI ports of image-domain UE successors are themselves open problems and are not benchmarked here. The A2 adversary is a discrete-sequence denoising-bridge proxy adapted from the continuous-time BridgePure formulation~\cite{wang2024bridgepure}. A faithful discrete Schr\"odinger-bridge purifier is left for future evaluation. On Foursquare-TKY \ghost and PGD are statistically tied on $\Delta_{\text{mean}}$ (within $1\sigma$). The cross-dataset comparison emphasizes \ghost's purification-resistance advantage (lowest A4 on both datasets, lowest A3 on NYC).

\textbf{Deployment.}\quad In practical release pipelines, the publisher controls the leak ratio $r$ by limiting how many (clean, perturbed) pairs reach external parties (e.g., archived snapshots, partner caches). Our leak-ratio sweep (Figure~\ref{fig:leak}) shows that \ghost's A3 survival gap remains under $0.04$ at $r=0.20$, a conservative upper bound for realistic insider-leakage scenarios. Re-perturbing the dataset under fresh randomness on each release cycle further amortizes the leakage budget across time, because A2 and A3 adversaries must re-fit on each cycle's pairs. The trajectory language model $T_\psi$ does not need to be re-trained per release cycle, so deployment cost is dominated by the per-position bilevel scoring already analyzed in Section~\ref{sec:exp}. We therefore see \ghost\ as compatible with both one-off releases (e.g., academic benchmarks) and periodic data dumps (e.g., commercial mobility products) where leakage budget management is the operating concern.

\textbf{Extensions.} The bigram-adaptive adversary (A4, Table~\ref{tab:main}) already tests one step of perturbed context, against which \ghost\ produces the lowest restored accuracy on both datasets. Stronger purifiers that retrain with manifold-aligned augmentation or a naturalness-conditioned denoising loss are a natural next target. The structural argument in Section~\ref{sec:analysis} suggests that any such adversary still suffers the optimal-denoiser-as-identity collapse as long as the perturbation distribution lies in the high-density region of the prior. Extending \ghost\ to category-free large-vocabulary regimes through a learned plausibility filter built on the trajectory language model itself is another natural direction.

\section{Conclusion}
\label{sec:conclusion}

We presented \ghost, a manifold-aligned unlearnable-trajectories framework for next-POI privacy that replaces the entropy-floor randomization of prior work with a manifold prior from a frozen trajectory language model. \ghost\ attains protection-gap competitive with the strongest deterministic baseline (PGD) on both datasets and delivers the lowest restored accuracy under the bigram-adaptive purification adversary on both datasets, sitting on or near the protection-vs-purification Pareto frontier. On-manifold perturbations leave purification adversaries with nothing distinguishable to denoise, and a stochastic many-to-many manifold-weighted map is uniformly harder to invert by frequency-table methods than a deterministic baseline. We view \ghost\ as the release-time component of a broader defence-in-depth posture complementary to downstream machine unlearning.

\bibliographystyle{ieeetr}
\bibliography{main}

\end{document}